\documentclass[onecolumn,unpublished,a4paper]{quantumarticle}
\usepackage[utf8]{inputenc}
\usepackage[backend=biber,style=alphabetic,maxbibnames=999]{biblatex}
\usepackage{amsmath}
\usepackage{amssymb}
\usepackage{amsthm}
\usepackage{amsfonts}
\usepackage[caption=false]{subfig}
\usepackage[colorlinks]{hyperref}
\usepackage[all]{hypcap}
\usepackage{tikz}
\usepackage{relsize}
\usepackage{color,soul}
\usepackage{capt-of}
\usepackage{mathtools}
\usepackage{float}
\usepackage[section]{placeins}
\usepackage{listings}
\usepackage[T1]{fontenc}  
\usetikzlibrary{decorations.pathreplacing}
\usepackage{braket}

\DeclareFixedFont{\ttb}{T1}{txtt}{bx}{n}{4}
\DeclareFixedFont{\ttm}{T1}{txtt}{m}{n}{4}
\definecolor{deepblue}{rgb}{0,0,0.5}
\definecolor{deepred}{rgb}{0.6,0,0}
\definecolor{deepgreen}{rgb}{0,0.5,0}
\newcommand\cppstyle{\lstset{
language=C++,
basicstyle=\ttm,
otherkeywords={uint8_t, __m256i, size_t, ASSERT_TRUE, EXPECT_TRUE, TEST, BENCHMARK},
keywordstyle=\ttb\color{deepblue},
emphstyle=\ttb\color{deepblue},
stringstyle=\color{deepgreen},
commentstyle=\fontfamily{txtt}\selectfont\color{gray},
showstringspaces=false,
literate={*}{{\char42}}1
         {-}{{\char45}}1
}}
\lstnewenvironment{cpp}[1][]
{\cppstyle\lstset{#1}}{}

\newcommand\pythonstyle{\lstset{
language=python,
basicstyle=\ttm,
morekeywords={assert,as,echo},
keywordstyle=\ttb\color{deepblue},
emphstyle=\ttb\color{deepblue},
stringstyle=\color{deepgreen},
commentstyle=\fontfamily{txtt}\selectfont\color{gray},
showstringspaces=false,
literate={*}{{\char42}}1
         {-}{{\char45}}1
}}
\lstnewenvironment{python}[1][]
{\pythonstyle\lstset{#1}}{}

\lstdefinestyle{stimcircuit}{
    language=python,
    basicstyle=\fontsize{6}{6}\selectfont\ttfamily,
    upquote=true,
    stepnumber=1,
    numbersep=8pt,
    showstringspaces=false,
    breaklines=true,
    frame=single,
    aboveskip=1.5em,
    belowskip=1.5em,
    commentstyle=\color{gray},
    classoffset=1,
    morekeywords={DETECTOR,OBSERVABLE_INCLUDE,rec},
    keywordstyle=\color{deepgreen},
    classoffset=2,
    morekeywords={H,R,MPP,M,RX,RY,MY,MX,SQRT\_X,XCY,XCZ,YCX},
    keywordstyle=\color{deepblue},
    classoffset=3,
    morekeywords={X_ERROR,DEPOLARIZE2,DEPOLARIZE1},
    keywordstyle=\color{red},
    classoffset=4,
    morekeywords={TICK,SHIFT_COORDS,QUBIT_COORDS},
    keywordstyle=\color{gray}
}

\renewcommand\thesection{\arabic{section}}

\theoremstyle{definition}

\theoremstyle{definition}

\theoremstyle{definition}

\newcommand{\eq}[1]{\hyperref[eq:#1]{Equation~\ref*{eq:#1}}}
\renewcommand{\sec}[1]{\hyperref[sec:#1]{Section~\ref*{sec:#1}}}
\DeclareRobustCommand{\app}[1]{\hyperref[app:#1]{Appendix~\ref*{app:#1}}}
\newcommand{\fig}[1]{\hyperref[fig:#1]{Figure~\ref*{fig:#1}}}
\newcommand{\tbl}[1]{\hyperref[tbl:#1]{Table~\ref*{tbl:#1}}}
\newcommand{\theoremref}[1]{\hyperref[theorem:#1]{Theorem~\ref*{theorem:#1}}}
\newcommand{\definitionref}[1]{\hyperref[definition:#1]{Definition~\ref*{definition:#1}}}

\begin{document}
\title{Cleaner magic states with hook injection}

\date{\today}
\author{Craig Gidney}
\email{craig.gidney@gmail.com}
\affiliation{Google Quantum AI, Santa Barbara, California 93117, USA}

\begin{abstract}
In this paper, I show how an intentional hook error mechanism can be used as a control knob for injecting magic states into surface codes.
The limitation, and benefit, of this approach is that it can only inject states in the XY or YZ plane of the Bloch sphere.
This increases fidelity, because perturbations out of the target plane can be detected as errors.
I use Monte Carlo sampling to show that this technique outperforms previous injection techniques, achieving lower error rates at smaller spacetime cost under digitized circuit noise.
\end{abstract}

\emph{The source code that was written, the exact noisy circuits that were sampled, and the statistics that were collected as part of this paper are available at \href{https://doi.org/10.5281/zenodo.7575030}{doi.org/10.5281/zenodo.7575030}~\cite{gidneyhookdata2023}.}

\maketitle

\section{Introduction}
\label{sec:introduction}

The surface code is a leading candidate for use in large scale fault tolerant quantum computers, because of its high threshold and low connectivity requirements~\cite{fowler2012surfacecodereview}.
Although the surface code is lenient on quality, it is demanding on quantity.
The spacetime cost of non-Clifford operations, such as T gates and Toffoli gates, is particularly large.
The most efficient known technique for performing these gates is magic state distillation, which is expected to cost millions or tens of millions of qubit·rounds per non-Clifford gate~\cite{litinski2019lesscostly,gidney2019catalyzeddistillation,fowler2018latticesurgery}.

Magic state distillation works by using partially destructive cross-checks to transform noisy input states into a smaller number of less noisy output states~\cite{bravyi2005distillation}.
A key factor in the cost of magic state distillation is the injection error rate; the error rate of the initial magic states before distillation.
For example, when using the $15T \resizebox{!}{8pt}{$\xrightarrow{35p^3}$\,} T$ factory~\cite{bravyi2005distillation}, improving the error rate of injected states from $0.1\%$ to $0.01\%$ allows classically intractable algorithms such as \cite{lee2021hypercontraction,gidney2021factor,soeken2020improved} to be performed with one level of distillation instead of two.
Reducing the number of levels from two to one would have a large impact on the overall cost of non-Clifford gates, because it reduces the number of T state injections per logical T gate from 225 to 15, and per Toffoli gate from 900 to 60.
Even accounting for varying code distances within the factories distilling these injected states, this would cut the spacetime cost of non-Clifford gates by a factor of 3.
More generally, having a better injection error rate creates slack throughout the distillation process.
This slack can then be used to reduce costs in a variety of ways.
For example, an injection error rate that isn't low enough to omit the second $15T \resizebox{!}{8pt}{$\xrightarrow{35p^3}$\,} T$ factory may still be good enough to replace that factory with an $8T \resizebox{!}{8pt}{$\xrightarrow{28p^2}$\,} CCZ$ factory~\cite{jones2013}, reducing the number of T state injections per logical Toffoli gate from 900 to 120~\cite{gidney2019catalyzeddistillation}.
Alternatively, the code distance used during the first level of distillation could be reduced.

\begin{figure}
    \centering
    \resizebox{0.5\linewidth}{!}{
    \includegraphics{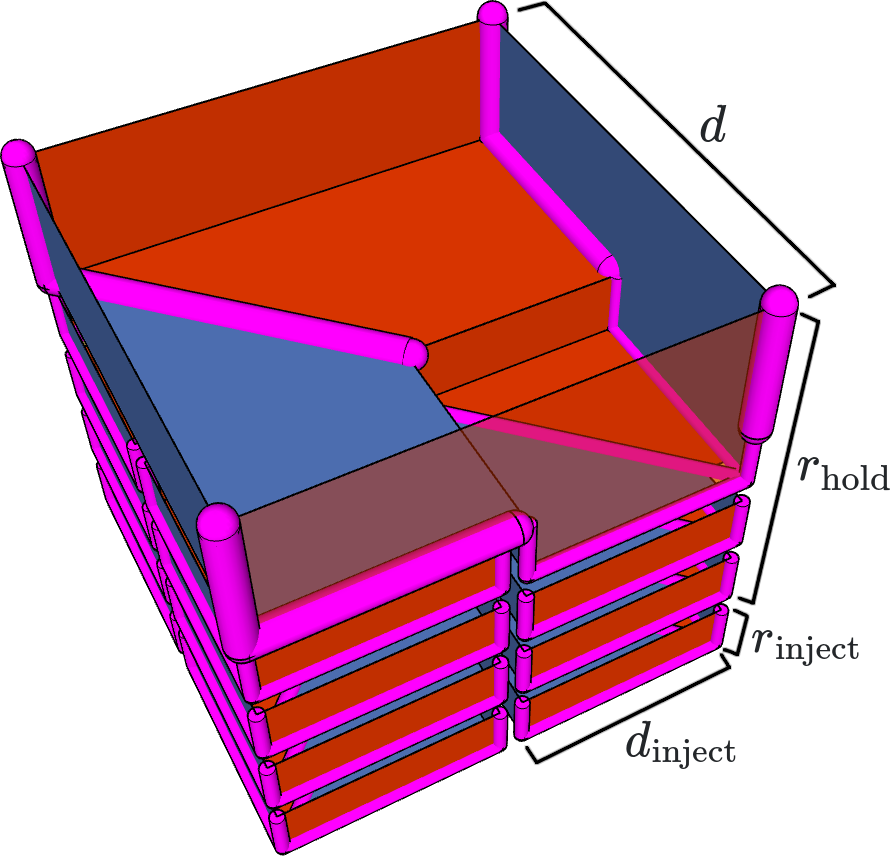}
    }
    \caption{
        Defect diagram of the injection process.
        Red surfaces are X boundaries, blue surfaces are Z boundaries, and magenta bars are twist defects.
        Attempts are made in parallel at a low distance $d_\text{inject}$, until an attempt succeeds and expands to the target distance $d$, as in \cite{li2015injection,singh2022injection}.
        Each attempt takes $r_\text{inject}$ rounds.
        Assuming there is some fixed deadline by which the injected state is needed, the state may need to be held idle for up to $r_\text{hold}$ rounds after it expands.
    }
    \label{fig:model}
\end{figure}

In this paper, I compare to two previous magic state injection methods: \cite{li2015injection} and \cite{singh2022injection}.
In \cite{li2015injection}, magic states are injected by treating a physical qubit as a degenerate 1x1 surface code, expanding that surface code to an intermediate distance $d_\text{inject}$, staying there for a small number of rounds $r_\text{inject}$, and then finally expanding to the target distance.
If any detection events occur before the final expansion, the whole process is restarted.
Restarting at any sign of trouble decreases the error rate, and using a smaller patch during verification allows multiple attempts to be run in parallel (see \fig{model}).
\cite{singh2022injection} also uses an approach where verification is done at a smaller size, but improves the injection error rate by starting with a surface code patch that's already at distance 2.
The logical observable of this patch is rotated by using a $ZZ(\theta)$ operation between two of the data qubits.
This cuts down on the number of distance 1 error mechanisms present in the circuit.

This paper improves on these previous state injection methods in three ways.
First, I'll use the rotated surface code instead of the unrotated surface code.
This halves the number of physical qubits needed to reach a desired code distance.
Second, I'll use a different trick for starting from a distance 2 surface code patch: rotating the logical observable by introducing an intentional hook error.
This further reduces the number of distance 1 error mechanisms.
It also removes the need for hardware to implement a second type of two qubit interaction, and the need for additional data-qubit-to-data-qubit connectivity at the injection site.
Third, I'll polish the circuit used during the first round of stabilizer measurements, reducing the number of distance 2 error mechanisms.

The paper is organized as follows.
In \sec{construction}, I explain the ideas behind hook injection, provide exact details of how to perform it, and discuss its limiting error mechanisms.
In \sec{benchmarking}, I benchmark hook injection against previous injection techniques and show that hook injection achieves lower error rates with smaller expected spacetime cost.
In \sec{conclusion}, I summarize the results and discuss potential future improvements.
The paper also includes \app{noise_model}, which defines the noise model used in simulations.

\section{Construction}
\label{sec:construction}

The primary factor that limits the fidelity of magic state injection is the possibility of single physical errors creating undetectable logical errors.
To first order, the goal of an injection strategy is to simply minimize the number of ways this can happen.

In the surface code, a ``hook error" is a single physical error that occurs on the measurement qubit halfway through the measurement circuit cycle, and causes a set of detection events equivalent to two data errors at the end of the cycle.
When compiling the surface code into a circuit, it's important to pick an order of operations that avoids aligning these two data errors in a direction that would help logical errors cross the patch.

Normally, hook errors are thought of as a problem.
But, in the context of magic state injection, they're also an opportunity.
During the surface code cycle, the hook error associated with a four body stabilizer can only occur at \emph{one specific time} on \emph{one specific qubit} and in \emph{one specific basis}.
This high specificity makes hook errors good control knobs for injection, because a hook error can be introduced without introducing a lot of other error mechanisms.

The basic setup is as follows.
A distance 2 surface code patch will be created, and its stabilizers will be measured.
The order of operations will be intentionally wrong, so that the patch's four body stabilizer has a hook error that directly rotates the logical observable.
This error will be used as a control knob to inject a state.
Because the hook error is specific to one basis, this control knob can only rotate around one axis of the Bloch sphere.
This limits the set of injected states to the XY plane of the Bloch sphere (or the YZ plane, depending on the basis of the four body stabilizer) but allows perturbations of the rotation axis to be detected as errors.

The patch is grown to an intermediate distance $d_\text{inject}$ as quickly as possible.
By careful ordering of operations, this can actually happen as the hook injection is occurring, while still measuring all stabilizers and without using additional layers of two qubit operations.
By the end of the first round of stabilizer measurements, the distance 2 patch has already grown to distance $d_\text{inject}$.

\begin{figure}
    \centering
    \resizebox{\linewidth}{!}{
    \includegraphics{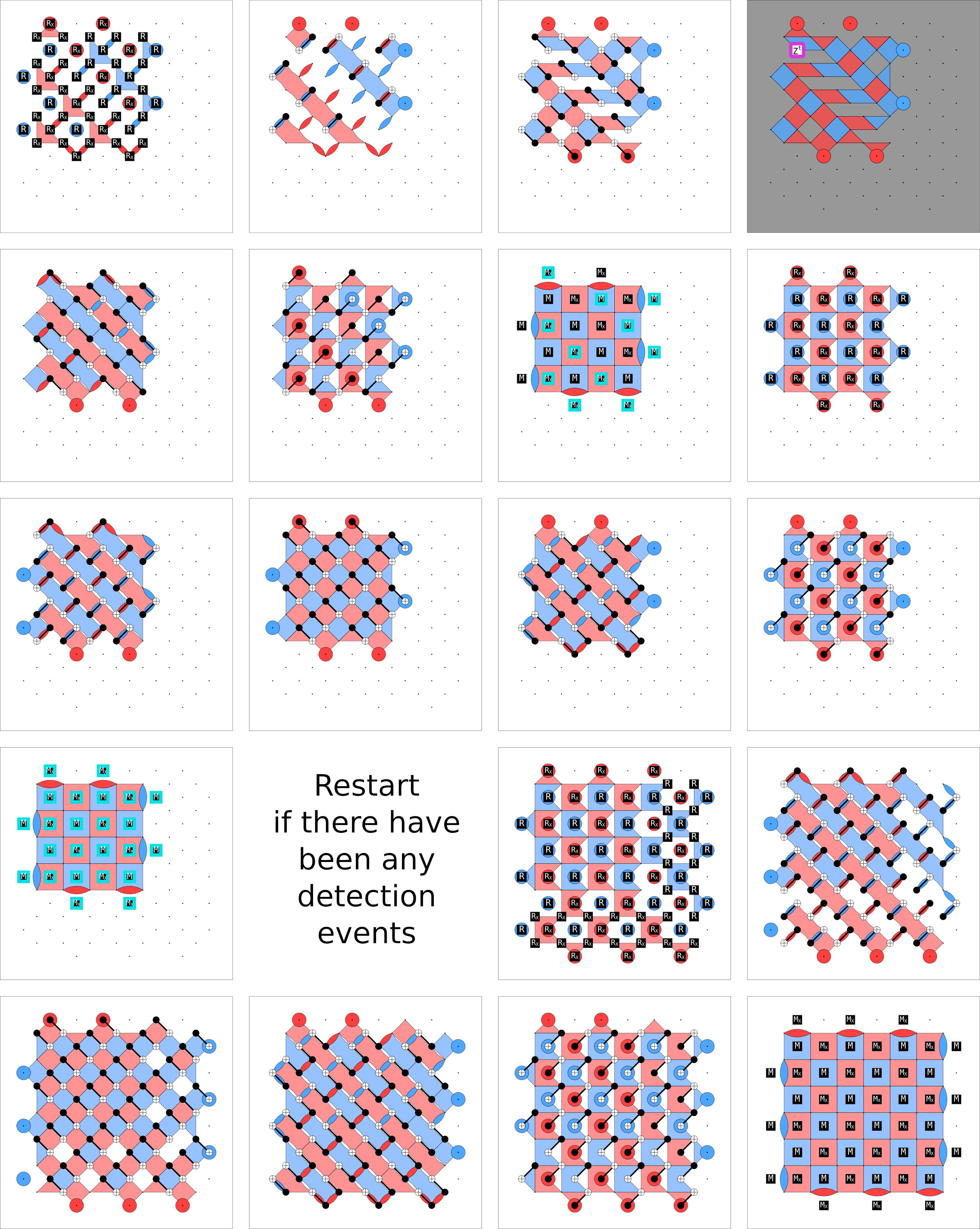}
    }
    \caption{
        Detector slice diagrams of hook injection with $d_\text{inject}=5$, $r_\text{inject}=2$, $d=7$.
        Layers advance in reading order, from left to right then top to bottom.
        Measurements outlined in teal are predictable from previous measurements, and are used to detect errors.
        If any of these measurements is incorrect, the injection process fails and restarts before expanding.
        The arbitrary-angle rotation, used to perform the injection, is the operation outlined in magenta during the shaded layer.
        Note that the exact gate order is important.
        For example, reversing the gate order of the second round would introduce additional distance 1 error mechanisms.
        Also note the first layer of CNOTs omits half of the CNOTs that are normally present, as they would have no effect on the initial state but would introduce additional error.
        Colored shapes correspond to stabilizers of the state that can be verified by comparing measurements.
        Uses the color convention RGB=XYZ (red is X, green is Y, blue is Z).
        See \fig{obs_slice} for corresponding observable slice diagrams.
    }
    \label{fig:construction}
\end{figure}

To minimize the parts of the patch that are at low code distance, the initialization basis of data qubits switches from X to Z across the patch's diagonal that crosses the injection site.
Additionally, all CNOT operations that would be no-ops (due to having a control qubit in the $|0\rangle$ state or a target qubit in the $|+\rangle$ state) are removed from the circuit.
This saves half the CNOTs in the first layer of two qubit gates, reducing noise.
The circuit resulting from these choices is the hook injection circuit, and is shown in \fig{construction}.

I view the fidelity of hook injection as being limited by three primary factors: (e1) distance 1 errors at the injection site, (e2) distance 2 errors at the injection site, and (e+) topological errors during the patch expansion.
There is also the possibility of topological errors while idling the state until it can be used, but generally this would be negligible due to being done at a code distance similar to the computation consuming the state.

Based on manual search and computer search, the compiled-to-CZ hook injection circuit has four digitized distance 1 error mechanisms:

\begin{itemize}
    \item 1 term in the DEPOLARIZE1 channel (has 3 terms) during the hook rotation.
    \item 2 terms in the DEPOLARIZE2 channel (has 15 terms) on the CZ before the hook error.
    \item 1 term in the DEPOLARIZE2 channel (has 15 terms) on the CZ after the hook error.
\end{itemize}

When injecting an $|i\rangle$ state, all four mechanisms cause a logical error.
Assuming a two qubit depolarization strength of $p$ and a single qubit depolarization strength of $p/10$, this places a hard lower bound of $7p/30$ on the injection error rate when injecting $|i\rangle$, no matter the amount of postselection.
This bound depends on the injected state.
For example, the $|+\rangle$ state is immune to one of the mechanisms and so has a more forgiving lower bound of $5p/30$.

A computer search of the compiled-to-CZ hook injection circuit found 200 digitized error mechanisms that are part of a distance 2 pair when injecting the $|+\rangle$ state, or 213 when injecting the $|i\rangle$ state.
Adding up their contributions, accounting for each error having a specific number of partners that complete it and the fact that depolarizing errors spread probability over multiple cases, gives an overall contribution of approximately $56p^2$ undetectable injection errors from distance 2 errors when injecting $|i\rangle$ and $21p^2$ when injecting $|+\rangle$.
The contribution to the injection error rate from distance 2 errors is roughly twice as large as the contribution from distance 1 error at $p=1\%$, five times smaller at $p=0.1\%$, and negligible at $p=0.01\%$.

Topological errors during the patch expansion are caused by error chains that partially cross the patch, while it's still at distance 2 or $d_\text{inject}$, resulting in an incorrect prediction by the decoder.
Most of these partial chains are removed by the postselection process, but some survive via timelike or spacetimelike errors pushing detection events past the last postselected round.
This kind of injection error can be suppressed by increasing $d_\text{inject}$ and $r_\text{inject}$.

Although the e1+e2 limit of hook injection is promisingly low, the benchmarking data shown in the next section makes clear that approaching this limit is expensive (e.g. discard rates above $99\%$).
At more reasonable discard rates, like $50\%$, and reasonable injection patch sizes, like $d_\text{inject}=5$, the injection error rate is a few times higher than the limit at a noise strength of $p=0.1\%$.

\begin{figure}
    \centering
    \resizebox{0.4\linewidth}{!}{
    \includegraphics{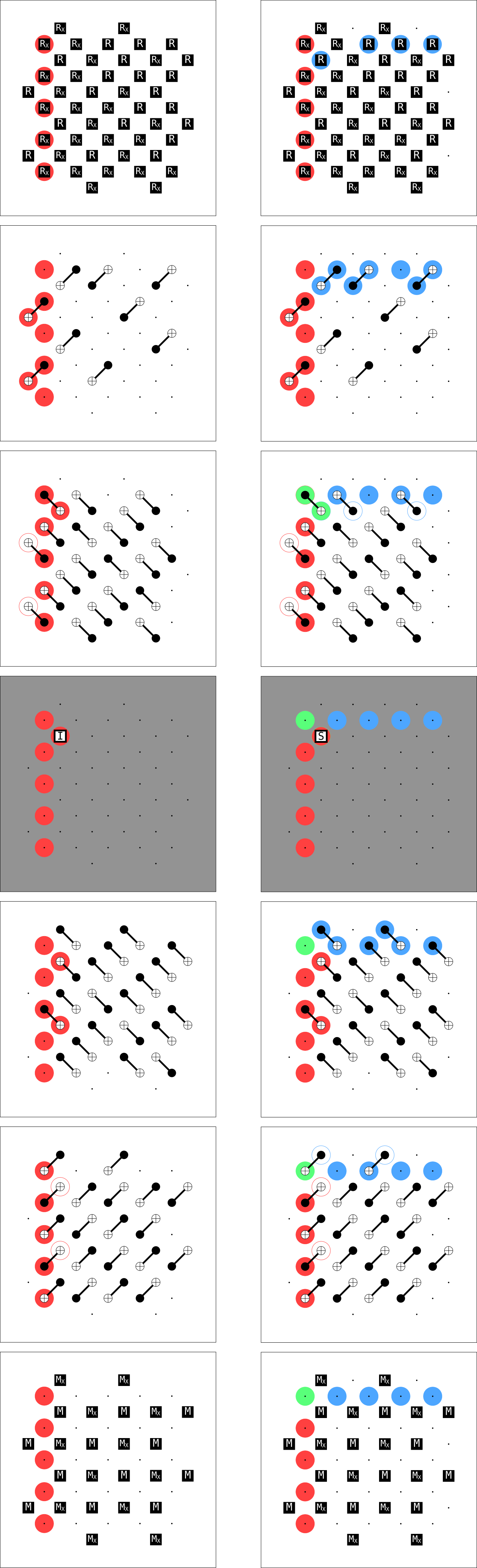}
    }
    \caption{
        Observable slice diagrams of the first round of injecting $|+\rangle$ (left) and $|i\rangle$ (right) using hook injection with $d_\text{inject}=5$, $r_\text{inject}=2$, $d=5$.
        The only distinction between the operations performed by the two circuits is the arbitrary-angle rotation in the shaded layer.
        Colored circles show the Pauli terms of the observable as the circuit progresses.
        Terms can be introduced by reset operations, are transformed by Clifford operations, and end up determining the observable whose eigenstate is the injected state.
        Filled in circles are the observable after the operations in a step have executed, while outlined circles are from before the operation.
        Uses the color convention RGB=XYZ (red is X, green is Y, blue is Z).
    }
    \label{fig:obs_slice}
\end{figure}

\section{Benchmarking}
\label{sec:benchmarking}

Before describing the simulations that I performed, I want to emphasize that I am \emph{not} claiming these simulations are definitively predictive of how magic state injection will perform on real hardware.
There are two major simplifications that I've made, which could cause the simulated behavior to differ from hardware behavior.

The first major simplification I'm making is assuming that noise is digitized.
I'm not simulating arbitrary quantum channels, I'm only simulating probabilistic Pauli channels.
This digitized noise approximation should be safest at high code distances, but the magic state injection constructions considered in this paper involve distance 1 errors where physical mechanisms directly perturb the logical state.

The second major simplification I'm making is that the states I'll simulate injecting are stabilizer states: $|i\rangle$ and $|+\rangle$.
This allows me to use existing tools to easily take lots of samples from circuits with large code distances.
I expect the injection error rate of any state on the XY plane of the Bloch sphere to be between the injection error rates of these two states, but it's possible to construct noise models where that isn't the case.
For example, consider a hypothetical system where the dominant error is drift around the $X+Y$ axis or else drift around the $X-Y$ axis. These errors Pauli twirl into the same digitized noise model, but differ in how they affect a T state.
An X+Y rotation causes no change to a T state at all, while an X-Y rotation maximally perturbs T states (even more than it perturbs $|+\rangle$ or $|i\rangle$ states).
So, although I do expect digitized simulations to be qualitatively predictive of hardware, I caution the reader that the exact numbers can't truly be trusted until they come from real experiments.

\begin{figure}
    \centering
    \resizebox{0.75\linewidth}{!}{
    \includegraphics{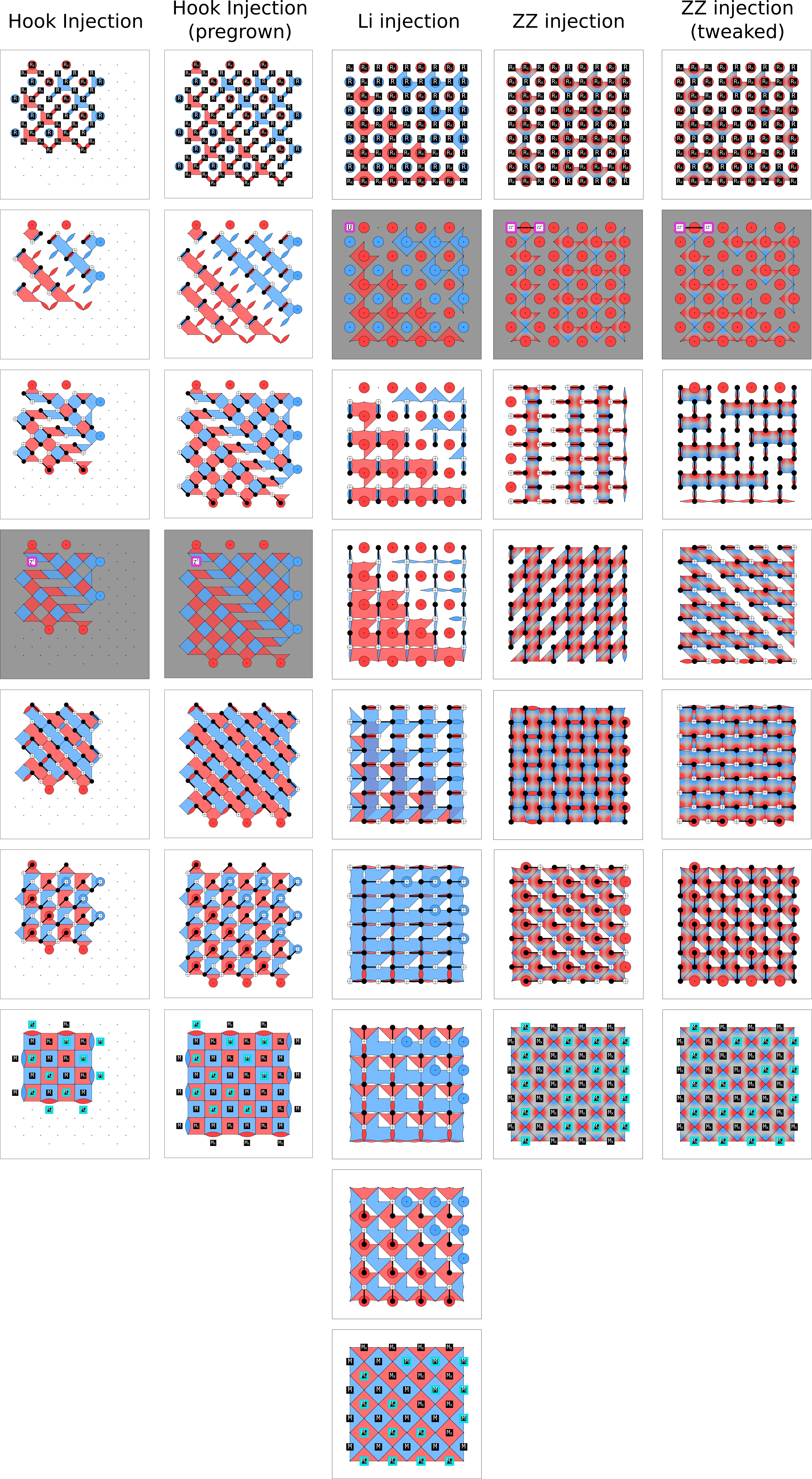}
    }
    \caption{
        Detector slice diagrams of the injection strategies compared in this paper, including Li injection~\cite{li2015injection} and ZZ injection~\cite{singh2022injection}.
        The displayed circuits are for $d_\text{inject}=5$ and $r_\text{inject} = 2$.
        Measurements outlined in teal are predictable and used for postselection.
        The pregrown hook injection differs from the others in that it only postselects a sub-region of the measurements.
        Each construction's injection operation is outlined in magenta in the layer shaded gray.
        The tweaked ZZ injection changes the initialization basis pattern and the order of operations to work better for unbiased noise.
        When benchmarking, these circuits were transpiled to use CZ gates.
    }
    \label{fig:construction_comparison}
\end{figure}

To compare hook injection to previous work, I reproduced the circuits from \cite{li2015injection} (called ``Li injection" in figures) and from \cite{singh2022injection} (called ``ZZ injection" in figures) as exactly as possible.
Because ZZ injection was built specifically for biased noise, but I am using unbiased noise, I also made a variant of ZZ injection with some details changed to improve its performance under unbiased noise (called ``ZZ (tweaked) injection" in figures).
I also tested a variant of hook injection where the patch started at the full target size, and the meaning of $d_\text{inject}$ was changed to be the diameter of the region of postselected measurements (called ``hook (pregrown) injection" in figures).
The circuit schedules of the first round of each of these constructions is shown in \fig{construction_comparison}.

The task I decided to compare the constructions on was preparing an $|i\rangle$ state with a target distance of $d=15$, with $r_\text{hold}=15$ rounds of idling after postselection ended.
I used the noise model in \app{noise_model}, with a noise strength of $p=0.1\%$.
At the end of these simulated injections, I used magically noiseless measurement of all stabilizers and relevant observables to check the state.
All circuits were transpiled to use CZ gates before being benchmarked.
Errors were decoded using an internal correlated minimum weight perfect matching decoder written by Austin Fowler.

Each circuit construction is parameterized by $d_\text{inject}$ and $r_\text{inject}$.
I simulated all variants satisfying $2 \leq d_\text{inject} \leq 7$ and $1 \leq r_\text{inject} \leq 6$.
I stopped at $d_\text{inject}=7$ because this was the maximum size that allowed parallel attempts with a target distance of 15.
Specifically for the pregrown hook injection construction, where this wasn't a relevant constraint, I increased the bounds to $2 \leq d_\text{inject} \leq 11$.
All variants were sampled until a billion samples had been taken, or 1000 undiscarded logical errors had occurred, whichever came first.

For each variant I combined the sampled discard rate, the injection patch size ($d_\text{inject}$), and the number of check rounds ($r_\text{inject}$), into an estimate of the expected spacetime volume cost per successful injection.
To highlight tradeoffs between spacetime cost and injection error rate, I plotted each construction's Pareto frontier in \fig{frontier}.
These frontiers clearly show that hook injection outperforms the previous constructions.
Hook injection reaches lower injection error rates than the previous constructions, and it achieves specific error rates with lower cost than the previous constructions.

\fig{frontier} can also be used to inform the choice of configuration parameters.
For example, I'm interested in creating injection processes that fit into larger factories, where the injection has to pack into a limited amount of spacetime.
In this context, the injection taking too long to succeed causes the factory to fail.
Adding more detail to the example, suppose the injection site is a distance 15 patch, available for 15 rounds, and the chance of the injection succeeding, before that deadline is reached, needs to be at least 99\%.
Running a distance 15 patch for 15 rounds costs around 7000 qubit·rounds.
The half life of a repeat-until-success process is roughly 70\% of its expected duration.
It takes 7 half lives to reach a 99\% chance of completion.
Thus, we are interested in injection configurations with an expected cost between 500 qubit·rounds and 2000 qubit·rounds.
The best choice, for this particular budget, would be hook injection with $d_\text{inject}=7,r_\text{inject}=2$ (injection error rate $0.06\%$, expected cost 700 qubit·rounds) or $d_\text{inject}=6,r_\text{inject}=3$ (injection error rate $0.04\%$, expected cost 950 qubit·rounds).
Alternatively, if I wanted a much higher success-before-deadline probability for the injection, I could pick $d_\text{inject}=5,r_\text{inject}=2$ (injection error rate $0.09\%$, expected cost 200 qubit·rounds).

\begin{figure}[h]
    \centering
    \resizebox{\linewidth}{!}{
    \includegraphics{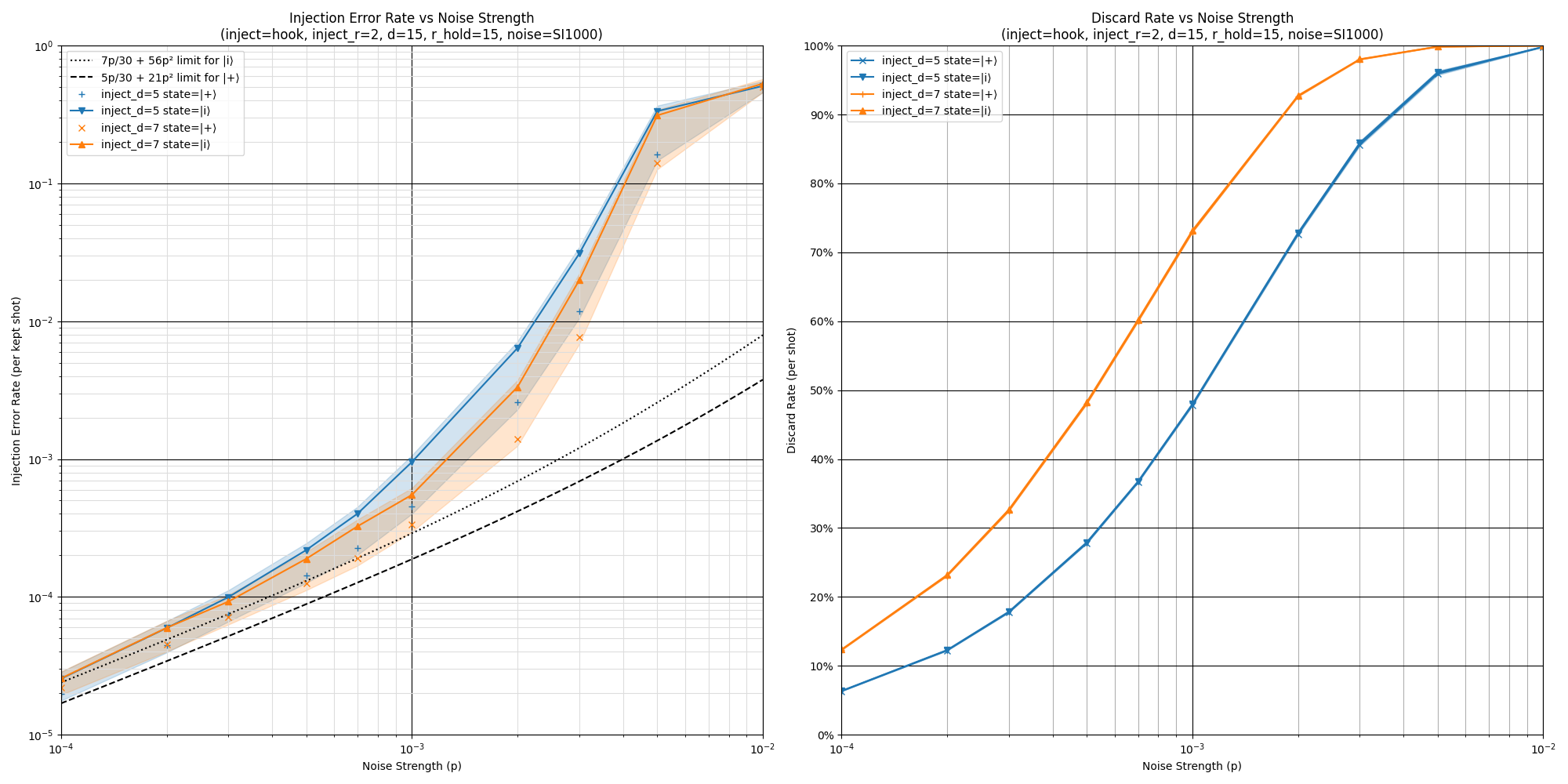}
    }
    \caption{
        Injection error rate and discard rate of hook injection, using 2 rounds of postselection and an injection distance of 5 or 7, as noise strength is varied.
        The $5p/30 + 56p^2$ and $7p/30 + 21p^2$ limits come from counting undetectable distance 1 and distance 2 errors in the hook injection circuit.
        The top of the shaded regions corresponds to the largest hypothesis probability with a likelihood within a factor of 1000 of the max likelihood hypothesis, given the data, when injecting $|i\rangle$.
        The bottom of the shaded regions corresponds to the smallest hypothesis probability with a likelihood within a factor of 1000 of the max likelihood hypothesis, given the data, when injecting $|+\rangle$.
        The two injected states have different injection error rates because of differences in what noise they are sensitive to (e.g. a $|+\rangle$ state isn't affected by an X error).
        Injecting a T state, instead of $|i\rangle$ or $|+\rangle$, should have an injection error rate near the middle of the corresponding shaded area.
    }
    \label{fig:expected_usage}
\end{figure}

Based on the above reasoning, I selected hook injection with $d_\text{inject}=5,r_\text{inject}=2$ and hook injection with 
$d_\text{inject}=7,r_\text{inject}=2$ as variants to more closely explore.
For these variants I simulated injecting $|i\rangle$ and also simulated injecting $|+\rangle$.
I also simulated a variety of noise strengths, ranging from $0.01\%$ to $1\%$.
I again used $d=r_\text{hold}=15$, but this time I didn't use magically noiseless measurement.
To check the injected $|+\rangle$ states I used transversal X basis measurement, and I used the inplace Y basis measurement from \cite{gidney2023ybasis} to check the injected $|i\rangle$ states.
Because of the lack of any magically noiseless steps, these injection experiment circuits could be run unmodified on real hardware.
The results of these simulations are shown in \fig{expected_usage}.
They show that hook injection achieves good injection error rates, with reasonable discard rates, over a large range of plausible physical error rates.
In particular, at $p=0.1\%$, the $d_\text{inject}=5$ variant achieves an injection error rate below $0.1\%$ and a discard rate below $50\%$ while the $d_\text{inject}=7$ variant achieves an injection error rate below $0.06\%$ and a discard rate below $75\%$.

\section{Conclusion}
\label{sec:conclusion}

In this paper, I described a new magic state injection technique based on introducing an intentional distance 2 hook error.
I benchmarked this technique, and showed it outperformed previous work.

\begin{figure}
    \centering
    \resizebox{\linewidth}{!}{
    \includegraphics{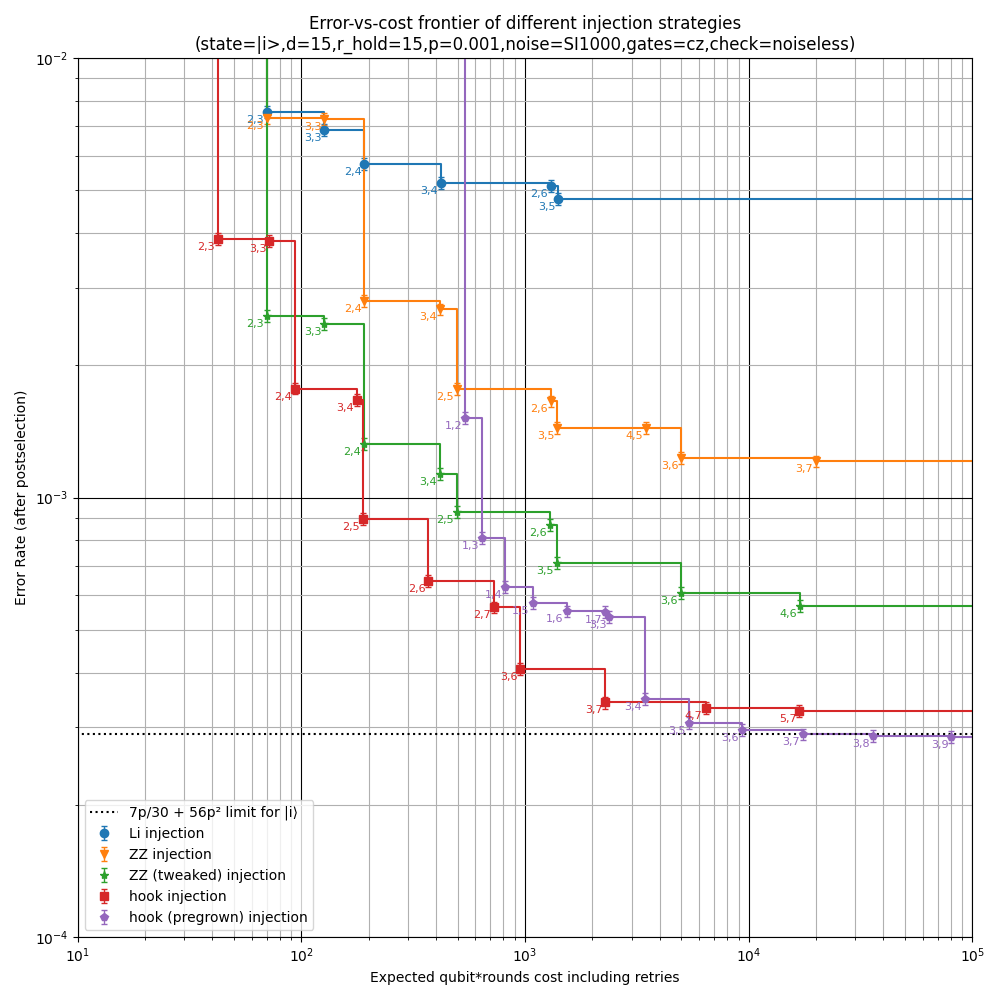}
    }
    \caption{
        The frontier of expected costs paid to achieve desired injection error rates, for different injection strategies.
        The small numbers next to each data point are the pair $(r_\text{inject},d_\text{inject})$, indicating how many rounds of postselection were used and what the distance was during the postselection process before the patch was expanded to the target distance of 15.
        The ``pregrown" strategy differs from the others in that it starts at the target distance, with the injection distance determining the region of measurements that are postselected instead of the initial patch size.
        Within each injection strategy, dominated points have been omitted.
        Each data point represents 1000 sampled errors or a billion attempts, whichever came first when sampling.
        Data points with discard rates so high that fewer than 100 sampled errors were seen have been omitted.
        Note that the right hand side extends comically far: $10^5$ qubit·rounds is roughly the cost of an entire first stage distillation factory~\cite{fowler2018latticesurgery,gidney2019catalyzeddistillation}.
        The $7p/30 + 56p^2$ limit comes from counting undetectable distance 1 and distance 2 errors in the hook injection circuit.
    }
    \label{fig:frontier}
\end{figure}

Although hook injection substantially improves over previous injection techniques, I want to emphasize that there are still many ways that magic state injection could be further improved.
For example, the performance of the pregrown hook injection variant in \fig{frontier} makes it clear that good error rates can be achieved while postselecting only a subset of the detectors within the surface code patch.
It should be possible to find better tradeoffs between injection error rates and discard rates by more carefully choosing which detectors to postselect.
Another interesting avenue might be to combine hook injection with ZZ injection, allowing the surface code patch to be started at distance 4 at the cost of requiring one non-local physical gate.
I also want to point readers to \cite{chamberland2020inject}, which uses a completely different approach to state injection (a transversal Clifford operation in the color code controlled by a cat state with flag qubits), to emphasize that the design space of injection extends beyond simply finding the best place to insert one magical rotation.

I'll end the paper with a small irony.
Given the large improvement over previous work shown in \fig{frontier}, it would be reasonable to expect hook injection to substantially reduce the projected cost of magic state distillation.
But there's a problem: in previous estimates of the cost of magic state distillation made by myself and Austin Fowler~\cite{fowler2018latticesurgery,gidney2019catalyzeddistillation}, we were already assuming these better injection error rates!
We failed to realize how much full circuit noise can hurt injection, compared to the interaction-dominated model emphasized in \cite{li2015injection}, even though that paper is clear on this point.
So, hook injection hasn't reduced the expected cost of magic state distillation.
Instead, hook injection has turned accidentally optimistic estimates into pragmatically justified estimates.

\section{Acknowledgements}

I thank Austin Fowler for reading drafts of this paper and providing comments that improved it.
I thank Hartmut Neven for creating an environment where this research was possible.

\printbibliography

\appendix

\section{Noise Model}
\label{app:noise_model}

\begin{figure}
    \centering
    \resizebox{\linewidth}{!}{
        \includegraphics{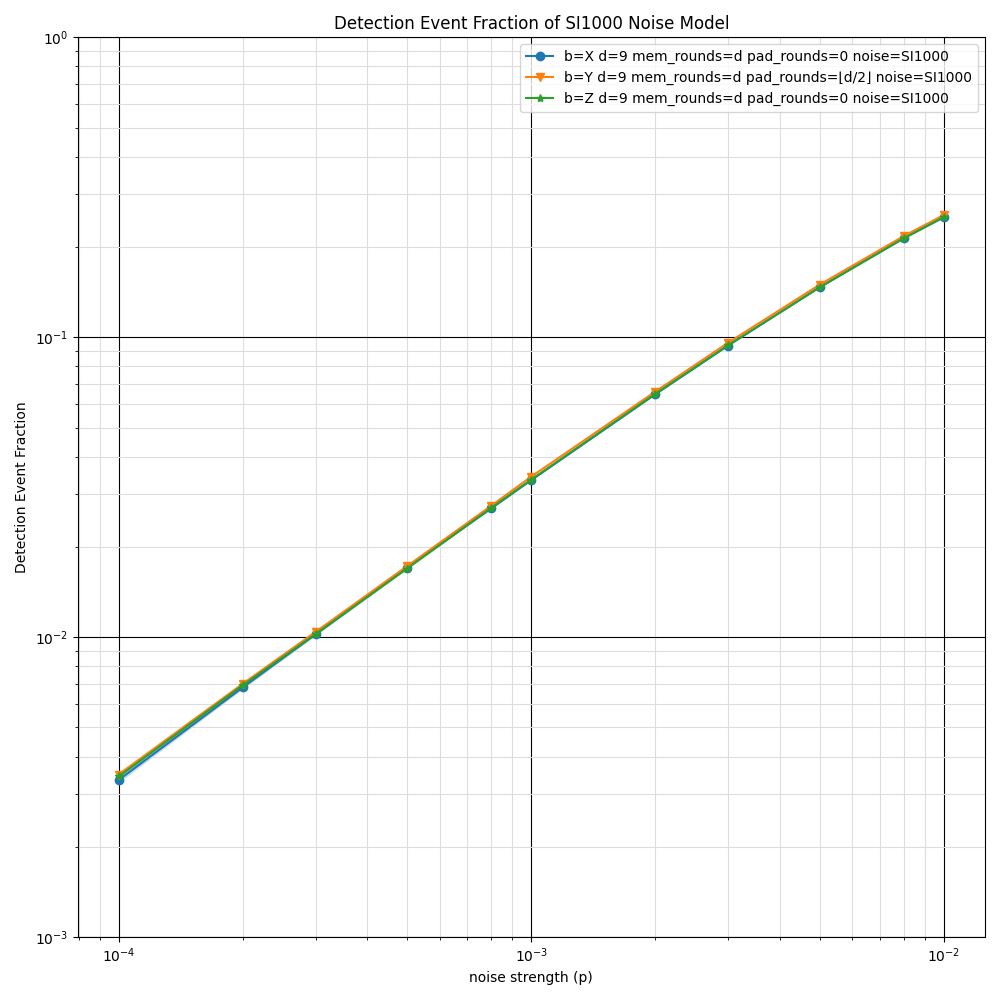}
    }
    \caption{
        Detection event fraction of basic memory experiment circuits in the SI1000 noise model.
        The detection event fraction is the probability of a detector producing a detection event, averaged over all detectors in the circuit.
    }
    \label{fig:det_frac}
\end{figure}

All circuits in this paper were simulated using the superconducting-inspired noise model defined in \tbl{noise_model}, called ``SI1000" (short for Superconducting Inspired with 1000 nanosecond cycle).
To provide a reference for comparisons to other error models or to hardware, the detection event fraction of the model at various noise strengths is plotted in \fig{det_frac}.

The source code at \cite{gidneyhookdata2023} can also use other noise models, in particular it includes a more uniform depolarizing model.
Hook injection also performs better under uniform depolarizing noise, but to keep the paper concise I have not included these simulation results.

\begin{table}[h]
    \centering
    \begin{tabular}{|r|l|}
    \hline
    Noise channel & Probability distribution of effects
    \\
    \hline
    $\text{MERR}(p)$ & $\begin{aligned}
        1-p &\rightarrow \text{(report previous measurement correctly)}
        \\
        p &\rightarrow \text{(report previous measurement incorrectly; flip its result)}
    \end{aligned}$
    \\
    \hline
    $\text{XERR}(p)$ & $\begin{aligned}
        1-p &\rightarrow I
        \\
        p &\rightarrow X
    \end{aligned}$
    \\
    \hline
    $\text{ZERR}(p)$ & $\begin{aligned}
        1-p &\rightarrow I
        \\
        p &\rightarrow Z
    \end{aligned}$
    \\
    \hline
    $\text{DEP1}(p)$ & $\begin{aligned}
        1-p &\rightarrow I
        \\
        p/3 &\rightarrow X
        \\
        p/3 &\rightarrow Y
        \\
        p/3 &\rightarrow Z
    \end{aligned}$
    \\
    \hline
    $\text{DEP2}(p)$ & $\begin{aligned}
        1-p &\rightarrow I \otimes I
        &\;\;
        p/15 &\rightarrow I \otimes X
        &\;\;
        p/15 &\rightarrow I \otimes Y
        &\;\;
        p/15 &\rightarrow I \otimes Z
        \\
        p/15 &\rightarrow X \otimes I
        &\;\;
        p/15 &\rightarrow X \otimes X
        &\;\;
        p/15 &\rightarrow X \otimes Y
        &\;\;
        p/15 &\rightarrow X \otimes Z
        \\
        p/15 &\rightarrow Y \otimes I
        &\;\;
        p/15 &\rightarrow Y \otimes X
        &\;\;
        p/15 &\rightarrow Y \otimes Y
        &\;\;
        p/15 &\rightarrow Y \otimes Z
        \\
        p/15 &\rightarrow Z \otimes I
        &\;\;
        p/15 &\rightarrow Z \otimes X
        &\;\;
        p/15 &\rightarrow Z \otimes Y
        &\;\;
        p/15 &\rightarrow Z \otimes Z
    \end{aligned}$
    \\
    \hline
    \end{tabular}
    \caption{
        Definitions of various noise channels.
        Used by \tbl{noise_model}.
    }
    \label{tbl:noise_channels}
\end{table}

\begin{table}
    \centering
    \begin{tabular}{|r|l|}
    \hline
    Ideal gate & Noisy gate
    \\
    \hline
    (any single qubit unitary, including idling) $U_1$ & $U_1 \circ \text{DEP1}(p / 10)$
    \\
    $CZ$ & $CZ \circ \text{DEP2}(p)$
    \\
    \hline
    $R_Z$ & $R_Z \circ \text{XERR}(2p)$
    \\
    $M_Z$ & $M_Z \circ \text{MERR}(5p) \circ \text{DEP1}(p)$
    \\
    \hline
    (Wait for $M_Z$ or $R_Z$) & $\text{DEP1}(2p)$
    \\
    \hline
    \end{tabular}
    \caption{
        The superconducting-inspired noise model used by simulations in this paper.
        Same as ``SI1000" from \cite{gidney2021honeycombmemory}.
        The single parameter $p$ sets the two qubit gate error rate, with other error rates being relative to this rate.
        Measurements are noisiest while single qubit gates are least noisy.
        Qubits not being reset or measured during layers containing measurements or resets incur additional depolarization on top of other error mechanisms.
        Note $A \circ B = B \cdot A$ means $B$ is applied after $A$.
        Noise channels are defined in \tbl{noise_channels}.
    }
    \label{tbl:noise_model}
\end{table}

\end{document}